\documentclass[12pt,preprint]{aastex}

\shorttitle{ }
\shortauthors{ }

\begin{document}

\title{ Eta Carinae's 2014.6 Spectroscopic Event: The Extraordinary   
   \ion{He}{2} and \ion{N}{2} Features\altaffilmark{1} }

\author{Kris Davidson\altaffilmark{2}, Andrea Mehner\altaffilmark{3},Roberta M.. Humphreys\altaffilmark{2}, John C. Martin\altaffilmark{4}, and Kazunori Ishibashi\altaffilmark{5}}

\altaffiltext{1} {Based on observations made with the NASA/ESA Hubble Space Telescope, 
which is opera ted by the Association of Universities for Research in Astronomy, Inc., 
under NASA contract NAS 5-26555.}
\altaffiltext{2} {Minnesota Institute for Astrophysics, 116 Church St SE, University of Minnesota, Minneapolis, MN 55455}
\altaffiltext{3} {ESO, Alonso de Cordova 3107, Vitacura, Santiago de Chile, Chile}
\altaffiltext{4} {Barber Observatory, University of Illinois, Springfield, IL, 62703}
\altaffiltext{5} {Division of Elementary Particle Physics and Astrophysics, Graduate School of Science, Nagoya University, Nagoya, 464-8602, Japan}

\begin{abstract}
Eta Carinae's spectroscopic events (periastron passages) in 2003, 2009, and 
2014 differed progressively.  
\ion{He}{2} $\lambda$4687 and nearby \ion{N}{2} multiplet 5 
have special significance because they respond to very soft X-rays 
and the ionizing UV radiation field (EUV).     
{\it HST/STIS observations in 2014 show dramatic increases 
in both features compared to the previous 2009.1 event.\/}   
These results appear 
very consistent with a progressive decline in the primary wind density, 
proposed years ago on other grounds.  If material falls onto the 
companion star near periastron, the accretion rate may now have 
become too low to suppress  the EUV. 
\end{abstract} 



\section{\ion{He}{2} $\lambda$4687 emission in $\eta$ Car's spectroscopic events}

The appearance of $\eta$ Car has evolved  rapidly 
in the past 15 years \citep{jcm06a,jcm10,am10b,am12}. 
The primary wind may be returning to its pre-eruption state 
\citep{kd12,rmh08}, but in any case the spectrum and brightness have 
changed much faster than before.  Extraordinary clues are provided by 
``spectroscopic events'' that occur near each periastron passage of 
the companion star in its 5.5-year orbit -- see many refs.\ in 
\citet{ci12}, \citet{hm12}, and \citet{kd12}.   
The events observed in 1998, 2003, and 2009 did not match 
each other \citep{kd05,ks09b,am11b}.   
Here we report major differences between the 2014.6 event and 
its predecessors, observed with the Space Telescope Imaging 
Spectrograph ({\it HST\/}/STIS).

The 2003.5 and 2009.1 spectroscopic  events seemed alike in their 
early stages, but their X-rays and exotic \ion{He}{2} emission 
differed after periastron.   
Broad \ion{He}{2} $\lambda$4687 emission occurred during 
the first half of each event \citep{sd04,jcm06b,am11b,teo12},  
even though neither star can ionize He$^+$ enough to produce 
this recombination line.  Martin et al.\  
noticed its anti-correlation with the 2-10 keV X-rays, and explored 
relevant physics. 
They concluded that a flood of 50-700 eV photons arose as the 
colliding-wind shocks became unstable near periastron,  
indirectly exciting \ion{He}{2} $\lambda$4687.   Those soft X-rays 
carry far more energy flux than the 2-10 keV photons, but cannot be 
observed except via \ion{He}{2} $\lambda$4687.

In 2009,  unlike 2003, definite $\lambda$4687 emission briefly reappeared 
after periastron.  (Concerning the 2003 record, see {\S}5 below.)  
The hard X-rays reappeared immediately thereafter, much 
earlier than in 1998 and 2003.   Most likely, $\lambda$4687 
emission signaled the re-formation of large-scale shock structure which 
produces hard X-rays.  These and other data appear  consistent 
with lower gas densities in 2009 compared to 2003  
\citep{ks09b,am10b,am11b,corc10}.

Thus, a qualitative extrapolation of the 1998--2003--2009 data 
suggested that $\eta$ Car's 2014.6 event might have a 
brighter  \ion{He}{2} $\lambda$4687 ``second flash.''   
Here we report {\it HST\/}/STIS observations which confirm that suspicion.
Equally significant, they also reveal unexpected strength in a nearby 
UV-excited \ion{N}{2} multiplet that was only barely visible in 
2003 and 2009.  


\section{Observations and Data Reduction}  

{\it HST\/}/STIS plays 
a unique role in our knowledge of $\eta$ Car.  
Unlike ground-based spectrographs, STIS can observe the central 
object without serious contamination by ejecta located 
150--500 mas from the star \citep{ham12,rem12,jcm06a}.  
It also provides unrivaled data homogeneity 
over the time period 1998--2014, immune to atmospheric 
effects -- though it was not operational during the 2009.1 event.  
Moreover, STIS allows UV coverage.  For all these reasons, 
{\it HST\/} data provide  necessary anchoring points for ground-based 
observations (see {\S}5 below).

The STIS/CCD observations reported here cover wavelength range 4562--4841 
{\AA} with a 100 mas slit width and an integration time 
of 25.5 s on each occasion.  The observing dates in 2014 are listed in 
Table 1.  The instrument setup was like that  described in our 
earlier papers, especially \citet{jcm06b}. Details can be found 
in the STScI and $\eta$~Car Treasury Program archives, where the data 
will become publicly available in 2015.\footnote{ 
   http://archive.stsci.edu/hst/  and  
   http://etacar.umn.edu/. }  
Additional wavelength ranges will be reported in later papers, 
but \ion{He}{2} $\lambda$4687 and the special \ion{N}{2} emission 
are most critical. 

We used the data reduction methods employed by 
\citet{jcm06b} and \citet{am10b}, 
developed for the HST $\eta$ Car Treasury Project in 
2002--2004 -- providing better spatial resolution than the 
standard STScI pipeline \citep{kd06}.  They allowed us to extract 
reliable spectra with spatial widths of 150 mas and  
slightly unsharp edges to minimize pixelization effects.  
At $\eta$ Car's distance, 150 mas corresponds to a projected size of 
about 340 AU, larger than the companion star's orbit but small 
enough to exclude the bright ejecta knots which contaminate all 
ground-based spectra of this object.  Spectral resolution was 
roughly 40 km s$^{-1}$, much narrower than the features discussed below.  
Since each relevant CCD pixel had many thousands of counts, and each 
spectral feature included many pixels, statistical errors are negligible 
compared to systematic errors.  With HST's excellent spatial resolution,  
systematic effects result mainly from the complexity 
of $\eta$ Car's spectrum, unrelated to the instrument.

Regarding ``phase'' in the 5.5-year orbital cycle, we use the  
definition adopted long ago for the $\eta$ Car Treasury Project:  
period = 2023.0 days exactly, and phase = 0 at MJD 56883.0  
(2014 August 14), MJD 54860.0 (2009 January 29), etc. 
Advantages of this system are outlined by  \citet{am11b}.     
We express $t$, time from the nearest zero phase, in days.  The time 
of periastron is unknown but most likely occurs near $t \sim -10$ d.

Quoted wavelengths are in vacuum, and Doppler velocities are heliocentric.


\section{\ion{He}{2} $\lambda$4687 in 2014 }  

These observations were intended to answer two obvious questions:
(1) Was there a substantial difference between the 2014.6 and 2009.1 
events concerning \ion{He}{2} $\lambda$4687?   
(2) If so, was the post-periastron emission stronger in 2014, 
extrapolating the 2003--2009 difference? 
Both answers prove to be ``yes.''

For comparison purposes, Figure 1 shows typical STIS spectra during
and after the 2003.5 event, and in September 2013. In addition, the bottom 
panel shows a Gemini/GMOS observation in 2009 \citep{am11b}.  It has 
better spatial resolution than most other ground-based data, but nevertheless 
shows major extraneous features due to outlying ejecta.  (Compare it to the 
top panel in Fig.\ 1 and the second panel in Fig.\ 2, obtained at 
similar orbital phase.) 

The critical range 4675--4694 {\AA} is marked by vertical lines in Fig.\ 1.  
STIS sampled this wavelength region on 15 occasions {\it outside the 
spectroscopic events\/} from 1998 to 2012, and none of them showed any 
sign of the $\lambda$4687 feature on the star -- see, e.g., Figs.\ 3--6 
in \citet{jcm06b}.     They resembled 2003.88 in Figure 1 and 2014.85 
in Figure 2, except that most other emission lines progressively 
weakened from 2000 to 2012 \citep{am10b,am12}. 
The 2013.70 spectrum in Figure 1 shows two noteworthy departures 
from the earlier record:  a \ion{N}{2} multiplet around 4620 {\AA} 
had become conspicuous in absorption ({\S}4 below), and a possible 
weak emission bump can be seen near 4687 {\AA}.

Figure 2 shows our recent STIS spectra.  Elevated fluxes near 
4687 {\AA} -- the shaded areas in the figure -- occur 
only during  spectroscopic events.

Results for the $\lambda$4687 emission strength appear in Figure 3.  
This is a contentious topic,  and the chief issue is whether the emission 
differed substantially between 2014.6 and 2009.1.   Therefore    
we employ two largely independent measures of the emission strength 
and they both support the same conclusion.    EW1 samples only the  
the ``obvious bump'' in the spectrum with a peak near 4687 {\AA} 
(e.g., top panel 
in Figs.\ 1 and 2);  this is easy to measure but it omits much of 
the flux.   EW2 includes a larger fraction of the emission but 
requires an accurate measure of the true continuum.

For EW1 we simply estimate a linear pseudo-continuum based 
on data within the narrow range 4670--4700 {\AA}, and calculate 
net equivalent width   in the usual way.  A dashed line in the top 
panel of Figure 1 shows an example.  EW1 amounts to a 
lower limit, because its pseudo-continuum 
level often lies well above the true continuum.  
EW1 in the 2009.1 and 2014.6 events is shown in Figure 3a, 
where small data points represent Gemini/GMOS observations in 
2009 \citep{am11b}, while heavier black dots are the STIS results 
for 2014.  Limit marks on each STIS data point are not standard 
error bars;  instead they indicate a range of $\pm$1.5\% in the local 
pseudo-continuum. Levels outside this range do not appear credible 
when plotted with the data.  Likely errors are difficult to quantify 
because they are mainly systematic, but the limit marks in Fig.\ 3 
might be regarded as informal equivalents to normal 1.5$\sigma$ or 
2$\sigma$ errors.

During the first half of each spectroscopic event, 
EW1 was noticeably larger in 2014 compared to 2009 (Fig.\ 3a).  
But the second half of the event, particularly $t \approx +17$ d,
showed a more dramatic difference:  the second flash in 2014 was 
roughly twice as strong as its predecessor in 2009.

Figures 2 and 3b show that EW1 is less than half of the story, 
because broader emission fills the region around 4687 {\AA} during 
an event.   \citet{jcm06b} showed that flux levels near 4605 and 
4745 {\AA} measured $\eta$ Car's continuum sufficiently well in 
the 1998--2004 STIS data.  When \ion{He}{2} $\lambda$4687 was not present 
(i.e., at any time outside an event), linear continua  $f_\lambda^c\,$  
fit the 4605, 
4685, and 4745 {\AA} levels very well;  e.g., 2003.88 in Fig.\ 1.   
The apparent slope was small, with 
    $ f_\lambda^c(4685 \, \mathrm{\AA}) \; \approx \; 
    0.99 \, f_\lambda^c(4745 \, \mathrm{\AA})$.\footnote{ 
   The precise {\it apparent\/} slope is slightly influenced by 
   focus effects \citep{kd06}; but what matters here is long-term 
   consistency, which has been excellent with STIS.}  
Estimating the continuum in this way, and integrating the emission between 
4675 and 4694 {\AA}, we find a second equivalent width EW2 which is   
often much larger than EW1.  \citet{jcm06b}, \citet{am11b}, and 
\citet{teo12} used this procedure.   Emission outside the 4675--4694 {\AA}  
interval is blended with strong unrelated spectral features.

In 2014, a set of unusual \ion{N}{2} lines disqualified the 
4605 {\AA} window for continuum sampling (Fig.\ 2 and {\S}4).  
We must therefore base the underlying continuum on just the 
4745 {\AA} window plus earlier experience and known features.  
Since  4687 {\AA} and 4745 {\AA} differ by only 1.2\% in wavelength, 
no likely process would alter the continuum ratio   
$ f_\lambda^c(4687 \mathrm{\AA}) \, / \, f_\lambda^c(4745 \mathrm{\AA})$   
by more than a fraction of a percent.  (4687 {\AA} is on the insensitive 
Rayleigh-Jeans side of $\eta$ Car's SED.) 

We estimate EW2 values in the following  way, favoring slightly 
higher continuum to avoid exaggerating the 2014.6 event. 
First, we assume that $f_\lambda^c(4745 \mathrm{\AA})$ is 1\% higher 
than the average of $f_\lambda$ across several {\AA} near 4745 {\AA}. 
This correction is motivated by two weak absorption lines near 4750 
and 4763 {\AA} (see below).  Then we adopt 
$ f_\lambda^c(4687 \mathrm{\AA}) \; = \; f_\lambda^c(4745 \mathrm{\AA})$,   
integrate  $\; f_\lambda - f_\lambda^c \;$  from 4675 to 4694 {\AA},   
and calculate equivalent width EW2 accordingly.  
Very likely the two weak absorption lines depress 
$f_\lambda^c(4745 \mathrm{\AA})$ by less than 1\%, and earlier 
experience suggests that $f_\lambda^c(4687 \mathrm{\AA})$ is about 1\%  
less than $f_\lambda^c(4745 \mathrm{\AA})$.  Consequently our EW2 values 
may be roughly 0.3 {\AA} too low, i.e., we err on the side of skepticism.

In terms of EW2, Figure 3b shows that {\it the second $\lambda$4687 flash 
in 2014 was more than twice as strong as in 2009.\/}    
The small horizontal bar above each data point shows 
the value if we decrease $f_\lambda^c(4687 \mathrm{\AA})$ by 1\%, 
arguably more likely as noted above.  

Various objections to our measurements can be proposed, but they are 
weak when examined.   
The temporal sampling is adequate to establish the brightening, 
since all data in 2009 showed a continuous rise and decline through 
about 45 days.  Our 2014 observations sampled the rise, the near-maximum, 
and the decline, and {\it all three\/}  were much brighter than  
corresponding phases in 2009.    
(If the true maximum was substantially higher, then the 
2009--2014 increase is even larger than we report.)   
We cannot prove that extra emission in 4675--4694 {\AA} 
(shaded in Fig.\ 2) was entirely due to \ion{He}{2};    
but {\it something\/} changed in EW2 between 2009 and 2014,  
and it correlated with EW1 which certainly represents \ion{He}{2} 
emission.   Errors in the continuum slope cannot be large enough 
to affect the issue, for physical reasons mentioned above.  No evident 
process can enhance the continuum in a wavelength 
range less than 100 {\AA} wide.    The narrow interval near 
4745 {\AA} is indeed the best continuum sample, because emission and 
absorption features are known to affect all other wavelengths between 
4550 and 4800 {\AA} during the spectroscopic event.  A Thomson-scattered    
wing of \ion{He}{1} $\lambda$4714 cannot account for much of EW2, 
because that would entail a detectable slope in $f_\lambda$ 
around 4687 {\AA};  and, also, the corresponding long-wavelength wing 
is not strong enough.   Altogether, EW2 is well justified and EW1 
behaved similarly.  Conceivably the values for 2009 were severely 
underestimated because of ground-based defects mentioned above, 
but that would not explain why the second flash around $t \sim +20$ d 
changed far more than the first flash around $t \sim -30$ d.

In summary, the combination of EW1 and EW2 in Figure 3 establishes 
a conspicuous difference between 
the 2014.6 and 2009.1 events.  Its physical significance 
will be noted in {\S}5 and in later papers.


\section{Special \ion{N}{2} features and the EUV }   

\ion{N}{2} multiplet 5 at 4603--4644 {\AA}, barely noticeable in 2009, 
became quite conspicuous in 2014;  compare especially the 2014.57 panel 
in Figure 2 to any spectrum in Figure 1.   
This development indicates a strengthened  EUV radiation 
field.  The multiplet's {\it lower\/} level has energy 18.5 eV, far too 
high for normal thermal excitation.  As explained by \citet{am11a}, 
it is populated in $\eta$ Car's primary wind by photons from the hot 
companion star. Consequently multiplet 5 can scatter ambient radiation, 
thus producing absorption and/or pseudo-emission lines, depending  
on viewing direction.

In late 2013 this set of features appeared mainly in absorption (2013.70 
in Fig.\ 1), but during the 2014.6 event they became mainly emission 
(Fig.\ 2).  Mehner et al.\ sketched the geometrical circumstances in 
their Figure 5.   In late 2013 the secondary star passed between us 
and the primary, except for the orbit inclination. Thus it excited 
\ion{N}{2} along our line of sight to the primary wind, causing 
absorption in multiplet 5.  In mid-2014 the secondary star moved to the 
far side of the primary, unfavorable for absorption but allowing 
apparent emission which was really light scattered in gas   
around the sides of the configuration.  By November 2014 the 
radiative excitation weakened because the size scale had increased.

Why were these features strong in 2014 but not in 1998, 2003, and 2009?   
\ion{N}{2} multiplet 5 requires photons with $h{\nu}  \gtrsim  14.5$ eV 
to create N$^+$, and $h{\nu} \approx$ 18.5 eV to excite it. 
Normally these  come from the hot secondary star.  
During the earlier spectroscopic events, its UV was effectively 
suppressed (\citealt{zan84}, and other refs.\ in \citealt{kd12}).   
{\it Evidently the 2014.6 event was characterized 
by a stronger EUV radiation field,\/} presumably from the secondary 
star.

The \ion{N}{2} velocities vary in much the same way as some other 
features \citep{am11a,am11b}.  We do not discuss their kinematics 
here, because their strength has a more immediate significance
as outlined above.


\section{Discussion}   

A wealth of evidence affirms the rapid and progressive changes in  
$\eta$ Car since 1998;  see \citet{am12} and many refs.\ therein.   
But this development has occasionally been disputed (e.g.\   
\citealt{ad10,teo12,mad13}).  
For \ion{He}{2} $\lambda$4687, \citet{teo14} report ``no significant 
changes between 2009.0 and 2014.6.''  Figure 3 contradicts 
that assessment, and the two events differed in other 
respects concerning the UV radiation field.  The most natural 
explanation continues to be a progressive decline of the primary 
wind density, see many references in \citet{kd12} and \citet{am12}.     
Since the photosphere is located in the wind,  decreasing densities 
favor smaller photospheric radii, higher radiation temperatures, and 
higher ionization states.

Whatever caused the ``extra emission'' around 4687 {\AA} (shaded areas 
in Fig.\ 2), it manifestly correlates in time and wavelength with the 
undoubted \ion{He}{2} $\lambda$4687 feature measured by EW1.  
In the absence of any other credible candidate, broadened \ion{He}{2}  
emission is the simplest interpretation and its physics appears 
reasonable \citep{jcm06b}.  Column densities near periastron \citep{ci12} 
are large enough to produce Thomson scattering, which broadens some 
of the $\lambda$4687 emission but also reflects much of it away from us.   
The $\lambda$4687 minimum around $t \sim -10$ d may conceivably have 
been a quasi-eclipse by gas near the primary star, and the ``second flash'' 
may have been stronger in 2014 merely because intervening densities 
were lower than in 2009.  Quantitative models will require elaborate 
simulations of colliding-wind shock instability, breakup, and 
reformation, and there are more adjustable parameters than truly 
independent observables.  The main 
point at present is that $\eta$ Car's successive spectroscopic events 
do indeed differ in a progressive way.

Subtle effects in ground-based data have occasionally led to 
misunderstandings.  For example,  \citet{teo12} stated that weak 
\ion{He}{2} $\lambda$4687 persisted between events, with a strength 
that would have been obvious in the STIS data.  In fact this was 
presumably emission from outlying ejecta at $r \sim$ 0.3 to 1 arcsec, 
as explained in {\S}4.2 of \citet{am11b} and {\S}6.3 of \citet{jcm06b}.
Teodoro et al.\  also stated that a second $\lambda$4687 flash occurred 
in 2003, based on two or three instances of  EW2 $\sim$ 0.6 {\AA} from 
a relatively modest instrument.  Between those observations, however, 
STIS showed EW2 $<$ 0.2 {\AA} at $t$ = +15 d.  
Teodoro et al.\ speculated that this was a ``bad datum,'' which is 
highly unlikely in view of the proven nature of the instrument 
and the multiple consistency of Fig.\ 5 in \citet{jcm06b}.  Since STIS 
was the more capable instrument, and hard X-rays did not reappear 
soon after that time, most likely there was no strong $\lambda$4687 
flash near $t \, \sim \, +15$ d in 2003.

We mention these examples because 
they illustrate why {\it excellent data with high spatial resolution 
are essential for this problem.\/}  The bottom panel in our Figure 1, 
and also Figure 2 in \citet{teo12}, show that even the best ground-based 
spectra of $\eta$ Car have obvious contamination by outlying narrow-line 
ejecta.  (Compare them to the STIS spectra.)  Data obtained with wider 
slits or inferior atmospheric conditions may be appreciably worse;  
and the relative brightness of the ejecta is changing with time 
\citep{am12}.  The extraneous features cast doubt on continuum  estimates 
and other sensitive measures that rely only on ground-based data.  
STIS, by contrast, has far better spatial resolution, excellent S/N, 
and good long-term stability;   see especially {\S}4.2 in \citet{am11b}.  
Moreover, its data are publicly available.

\ion{N}{2} multiplet 5 samples different physical parameters than either 
the helium lines or the low-excitation features.  It depends mainly on 
EUV photons with energies between 14 and 20 eV, supplied by the hot 
companion star \citep{am11a}.   At most times the secondary star 
has $T_\mathrm{eff} \; \sim \; 40{\,}000$ K, producing copious   
radiation at 14--20 eV   \citep{am10a}.     The \ion{N}{2} features arise in 
relatively normal parts of the primary wind, not the shocked regions.  
Their prominence in 2014 implies a substantial EUV radiation field.

Traditionally,  $\eta$ Car's spectroscopic events were defined by a 
temporary lack of ionizing UV \citep{zan84}.  Either (1) the hot secondary 
star moved inside the diffuse primary photosphere, and/or (2) it accreted  
primary-wind material near periastron, thereby reducing its $T_\mathrm{eff}$ 
\citep{sok03,sok05,sok07,ks09a,ks09b}.  
Both of these possibilities require ambient densities 
above some critical level.  Therefore, the unpredicted strength 
of \ion{N}{2} multiplet 5 in 2014 strongly suggests that gas densities 
were lower than they were in previous events,  consistent with 
a progressively decreasing primary wind (\citealt{am10b,am11b,kd12} 
and refs.\ therein).  In any case it signals another physical 
difference between the 2014.6 event and its predecessors.

Additional {\it HST\/}/STIS results on $\eta$ Car's 2014.6 event 
will be reported in subsequent papers.  

{\it Acknowledgements:}  
As always, we are grateful to Beth Periello and other STScI staff 
members for help in scheduling the intricate STIS observations. 
These data swere obtained in HST program GO 13377, whose P.I.\ is
A. Mehner.  We also thank Gary Ferland for good advice concerning 
line identifications and related tasks.


\begin{deluxetable}{cccccc}    
\tabletypesize{\scriptsize} 
\tablecaption{STIS Observations of 4561--4841 {\AA} in 2014} 
\tablewidth{0pt} 
\tablehead{ 
\colhead{Date} &    
\colhead{MJD} &     
\colhead{t\tablenotemark{a}} &      
\colhead{$f_\lambda(4745 \, \mathrm{\AA})$\tablenotemark{b} }  &  
\colhead{EW1\tablenotemark{c}} &    
\colhead{EW2\tablenotemark{c}}      
      \\
\colhead{}        &    
\colhead{(days)}  &    
\colhead{(days)}  &    
\colhead{}        &    
\colhead{(\AA)}   &    
\colhead{(\AA)}        
   }
\startdata
 2014-07-13  &  56851.2  &  -31.8  &  1.74  &  1.68  &  2.77   \\
 2014-07-30  &  56868.1  &  -14.9  &  1.99  &  0.01  &  0.69   \\
 2014-08-15  &  56884.3  &   +1.3  &  1.47  &  0.17  &  0.90   \\   
 2014-08-31  &  56900.4  &  +17.4  &  1.45  &  0.89  &  2.53   \\	
 2014-09-17  &  56917.0  &  +34.0  &  1.53  &  0.51  &  1.15   \\
 2014-11-09  &  56970.7  &  +87.7  &  1.54  &  0.02  &  0.01   \\  
\enddata 
\tablenotetext{a}{Days from the standard zeropoint at MJD 56883.0;   
                   phase = $t/(2023 \, \mathrm{d})$. }      
\tablenotetext{b}{Apparent continuum in units of $10^{-11}$ 
                   erg cm$^{-2}$ s$^{-1}$ {\AA}$^{-1}$, not corrected 
		   for slit throughput;  see text. }
 \tablenotetext{c}{EW1 and EW2 are different types of 
	          equivalent width for emission near 4687 {\AA},  
                  see text. }    
\label{tab:table1}  
\end{deluxetable}



\begin{figure}
\figurenum{1}
\epsscale{0.45}
\plotone{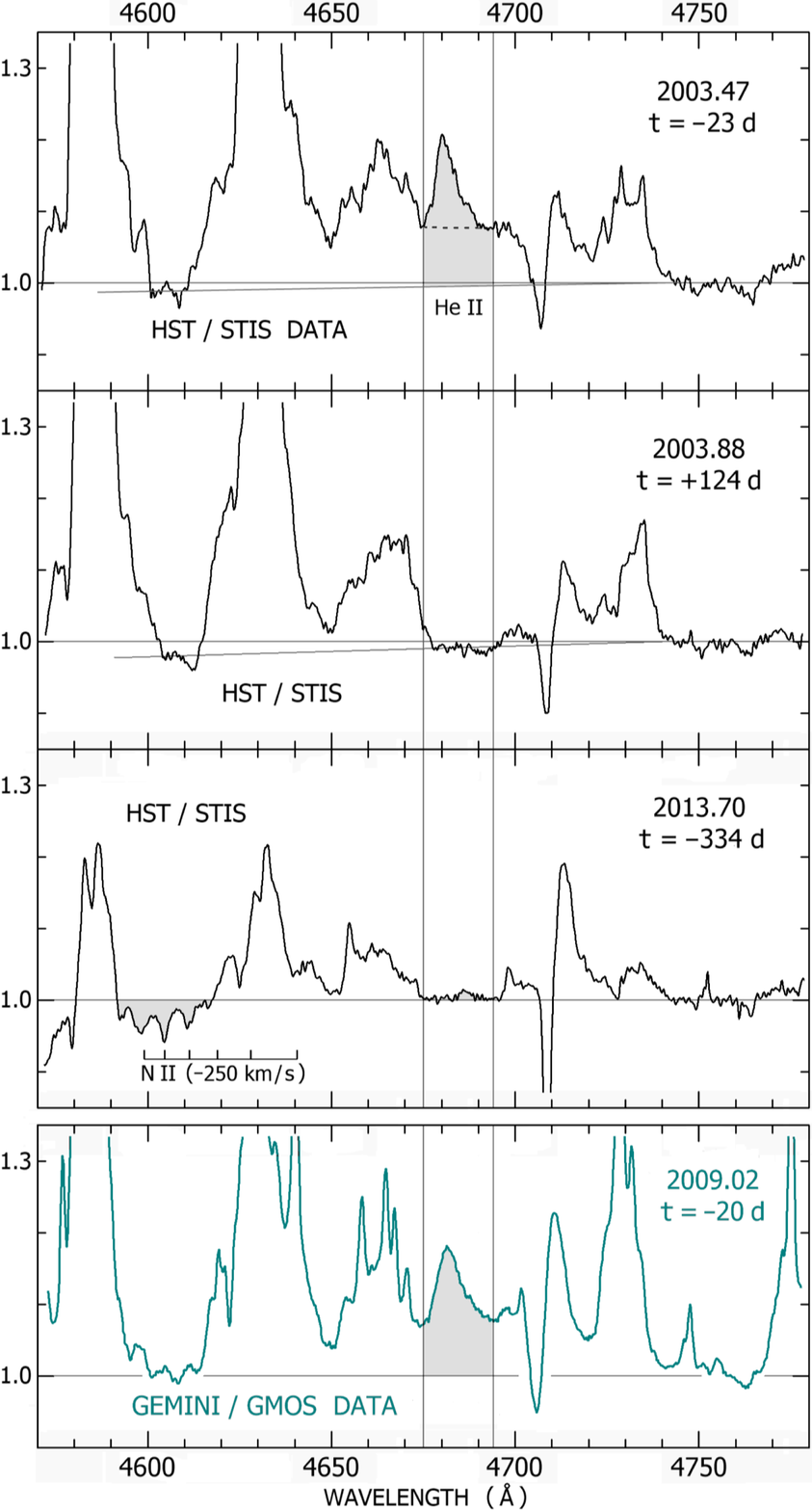}
\caption{Representative STIS data in 2003 and 2013, and ground-based 
   data in 2009.  Each tracing shows 
   $f_\lambda$ relative to the adopted value near 4745 {\AA}.  Baselines 
   with   small slopes in the 2003.47 and 2003.88 panels indicate 
   continuum fits.  A short dashed line in 
   the 2003.47 panel shows the base for a typical EW1 estimate, see text. 
   The bottom panel shows Gemini/GMOS data in the first 
   half of the 2009 event;  note the weakness of \ion{N}{2} features 
   around 4610 {\AA} compared to Fig.\ 2.  Like all other ground-based 
   spectra of $\eta$ Car, the Gemini data include many features 
   emitted by outlying ejecta. } 
\end{figure}


\begin{figure}
\figurenum{2}
\epsscale{0.45}
\plotone{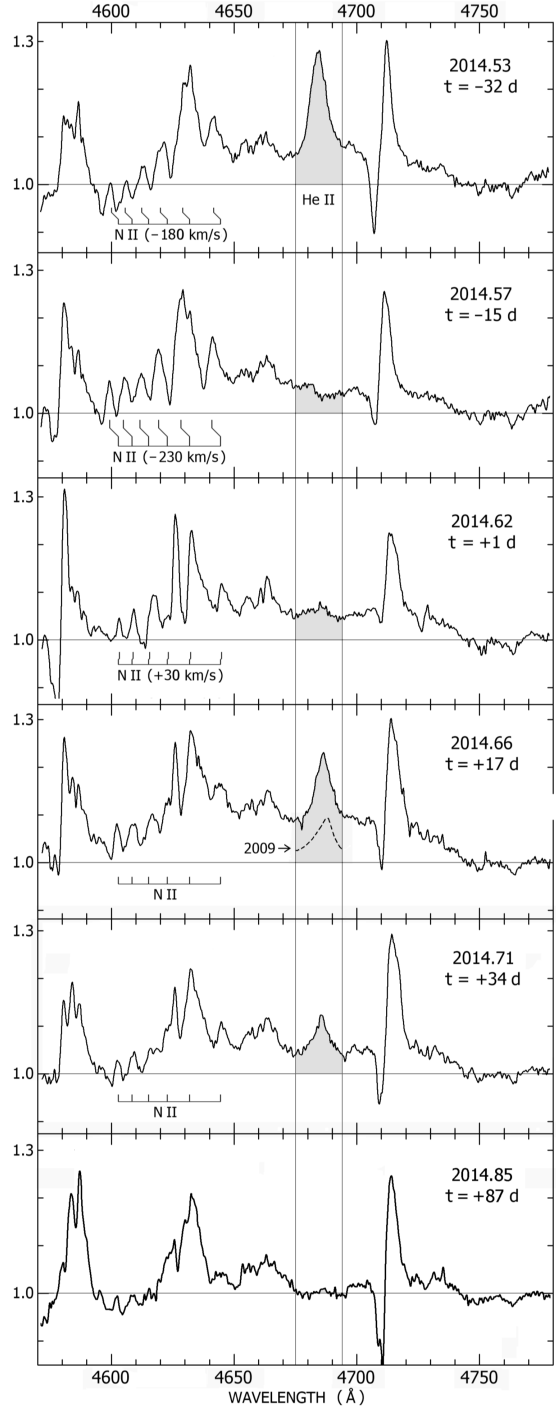}
\caption{STIS data in 2014, $f_\lambda$ relative to the adopted continuum
   value near 4745{\AA}. }  
\end{figure}

\begin{figure}
\figurenum{3}
\epsscale{0.55}
\plotone{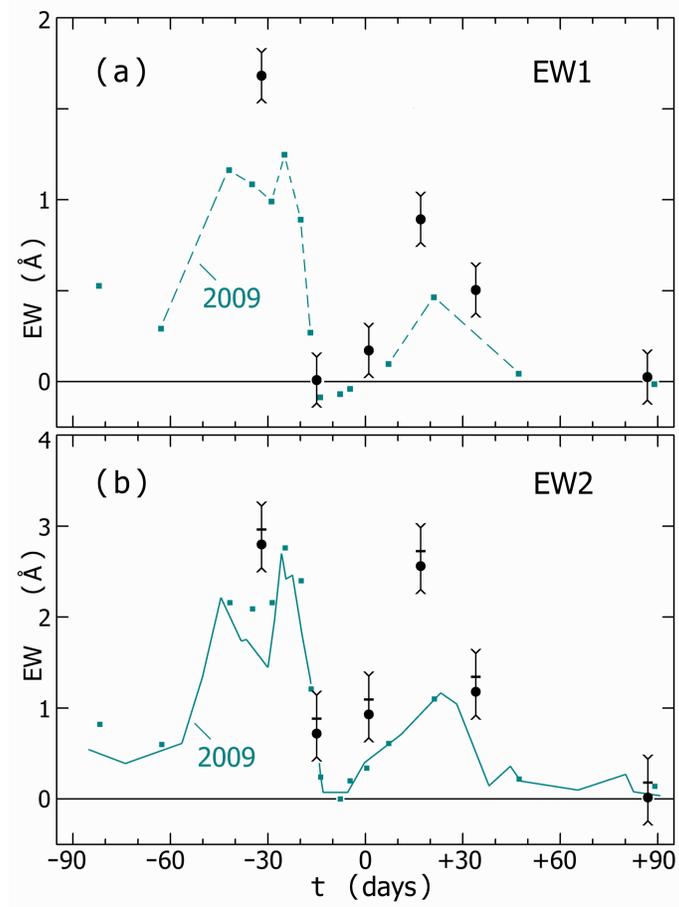}
\caption{ EW1 and EW2 for the emission near 4687 {\AA} in 2009 
  and 2014, see text.  The continuous curve in the lower panel 
  traces many data points from 2009 reported by   \citet{teo12}.  
  Small squares are other data points for 2009 from \citet{am11b}. 
  The heavy black data points show STIS results in 2014.  Their limit-marks 
  represent credible ranges of underlying continua, see text.  } 
\end{figure}  


\end{document}